# The Real Meaning of Quantum Mechanics

Francois-Igor Pris (И. Е. Присъ)

## Abstract

I suggest a contextual realist interpretation of Carlo Rovelli's relational quantum mechanics. The principal point is a correct understanding of the concept of reality and taking into account the categorical distinction between the ideal and the real. Within my interpretation, consciousness of the observer does not play any metaphysical role. The proposed approach can also be understood as a return to the Copenhagen interpretation of quantum mechanics, corrected within the framework of contextual realism. The contextual realism allows one to get rid of the metaphysical problems encountered by various interpretations of quantum mechanics, including the relational one. A wave function neither literally reflects an "exterior reality", nor is it merely a predictive mathematical tool. In particular, an entangled wave function can be viewed as the formal cause of quantum correlations whose reality is contextual. The reduction of a wave function is not a real physical process. And there are no autonomous quantum events which would be different from the facts of their representation. The quantum measurement problem results from a confusion between the categories of the ideal and the real. It is dissolved logically. The quantum theory plays the role of a Wittgensteinian rule (norm), "measuring" physical reality within a language game of its application.

1. **Introduction**

The entangled wave function is the cause of quantum correlations. The correlated quantum events are not autonomous, but are determined in the context of their observation. Independently from the means of observation, there are no events. The reduction of a wave function in a "process of measurement" is not a real physical process requiring an explanation, but a move into the context of measuring of a concrete value of a physical quantity. Respectively, measurement is not a physical interaction affecting the state of a system, but identification of a contextual physical reality. The measurement problem, or the problem of application of a physical theory to reality, appears as a result of confusion of the categories of the ideal (theory, concepts) and the real (application of the theory). It is dissolved logically. The problem is an instantiation of the Wittgenstein's rule-following problem. The quantum theory plays the role of a Wittgensteinian rule (norm), "measuring" physical reality within a language game of its application.

2. **Carlo Rovelli's interpretation of quantum mechanics**

Carlo Rovelli's relational interpretation of quantum mechanics is positioned by him as a realist in a weaker sense (albeit there are neo-Kantian, empiricist, and structuralist interpretations of his interpretation): a quantum state and values of quantum physical quantities are real only for (relatively to) an "observer" interacting with an observed quantum system while observing it. The role of the observer is played by any physical system interacting with the observed system [1, 2].

Rovelli's interpretation is closely related to the interpretations of quantum mechanics in the language of information theory. One can also establish connections between it and the "modal realism of substitution" of Michel Bitbol [3]. My own interpretation is a Wittgensteinian one within Jocelyn Benoist's "contextual realism" [4]. In many respects, it is close to the Bitbol's interpretation [3].[1]

## 3. Quantum Mechanics as a Norm for Measuring Reality

According to my interpretation, any theory, that is verified in experience and established, including quantum mechanics, has a domain of applicability, within which it is universal and, as a rule, gives truth and knowledge by its very nature. One can speak of the universal character of quantum mechanics only in this tautological or, better to say, logical sense. Beyond the domain of applicability of a theory, it is, strictly speaking, meaningless. Within the domain of its applicability an established theory acquires the status of a rule/norm for measuring reality, has a logical certainty. As a consequence, it is not falsifiable [6].[2]

An application of any theory, as well as any norm, rule or concept, including quantum mechanics, depends on a context. This is so in a broad sense – a theory is applicable only in the domain of its applicability (again, this is a tautology) – as well as in a narrow sense: any application of a theory within the domain of its applicability requires attention to the concrete conditions of its application (context). The notion of a context presupposes that of a norm, implicit or explicit in a given context. And vice versa: if there is a norm, there are its applications in context. As said, a theory plays the role of a norm (rule). Such a norm/rule is elaborated and anchored in reality. It is a norm/rule in the sense of the later Wittgenstein philosophy. That is why we speak of it as of a Wittgensteinian rule [6, 7].[3]

---

[1] Bitbol, for instance, speaks of the possibility to deduce a major part of the formalism of quantum mechanics from the two fundamental hypotheses: contextuality (or relativity) of determinations, and the invariance of the symbol, destined to predict their experimental values in different contexts. This also allows one to make quantum mechanics more comprehensible [5, p. 261].

[2] This does not mean that it cannot be "falsified" from the point of view of a more general theory, for example, quantum field theory. Similarly, Einstein's theory of relativity and quantum mechanics "falsify" Newton's mechanics, which, from their point of view (in their domain of applicability) turns out to be approximative or false (and, strictly speaking, meaningless) theory. At the same time, in the domain of its applicability the latter should be considered as a true theory, in the epistemic (empirical) sense as well as in the logical sense.

The problem of a rule/norm application in a context is the Wittgensteinian rule-following problem. The structure of this problem is that of the contextual realism. The quantum measurement problem as well as the hard problem in the philosophy of mind, the incompleteness of mathematics established by Gödel, and some other problems are instantiations of this Wittgensteinian problem. (Thus, there is a "family resemblance" between them.)

The gap between a rule (norm) and its application in the context of a "language game" or a "form of life" – is a logical gap, not a substantial one. Indeed, this is a gap between the category of the ideal and that of the real. The norms, rules, concepts belong to the former. Their applications belong to the latter [8, 9]. It is closed from the very beginning in a correct application of a Wittgensteinian rule, that is within an authentic language game (i.e. a language game having a justification post factum) [6, 10].

Our interpretation allows us to explain the EPR-paradox without invoking the hypothesis of non-locality of quantum mechanics. The «cause» (in a generalised sense) of quantum correlations is the entangled wave function. The correlated quantum events are not autonomous, but they are determined in the context of their observation. Independently from the means of their identification, there are no events. The reduction of a wave function in the «process of measurement» is not a real physical process, requiring an explanation, but a move to a context of measurement of a concrete value of a physical quantity. Respectively, the measurement is not a physical interaction leading to a change in the state of a system, but the identification of a contextual physical reality. That is, in a sense, in measuring (always in a context), one identifies just the fragment of reality where the (quantum) correlation takes place. As the elements of reality, the correlated events do not arise; they are. Only their identifications do arise.

Our interpretation allows us to demystify the Everett (or "many worlds") interpretation of quantum mechanics by contextualizing it, that is, by considering the Everettian worlds as all possible contexts. If the Everett interpretation is understood in the purely theoretical sense – as introducing a rule for measuring quantum reality – it is acceptable. However, a substantialisation of the Everettian rule entails a metaphysical many-world interpretation which is problematic [11]. In Kit Fine's terms, one could say that in this case the reality is fragmented [12]. From the metaphysical point of view, the fragmentation looks like the multiplicity of non-interacting ("parallel") worlds. However, in our view, it is more correct to say that reality is contextual.

---

[3] This is also in accord with Bitbol's view that quantum physics is a formalized know-how [21].

The measurement problem, or the problem of application of a quantum theory to reality, arises as a result of confusion between the category of the ideal and the category of the real. The theories belong to the former. Their applications belong to the latter. The problem is dissolved logically. It is, as said above, an instantiation of the Wittgensteinian rule-following problem. The role of a Wittgensteinian rule (norm) "measuring" physical reality within a language game of its application is played by the quantum theory.

## 4. A Contextual "Democratization" of the Copenhagen Interpretation

The Copenhagen interpretation of quantum mechanics is formulated in a language containing a reference to the observer: a quantum event does not just take place; it takes place for an "observer", whose role is played by a classical macroscopic system. Thus, one supposes the existence of a privileged system – a macroscopic observer not obeying the quantum laws.

In our view, there is a significant degree of truth here: an observer (if it is actually an observer) and what is observed belong to different categories. In doing so, the statuses «observer» and «observed system» are attributed in context.

Niels Bohr, already, introduced the context dependence of the epistemic (but not ontological) border between the observed system and the observer. In our view, a stronger statement can be made: the observer and the observed system can change their statuses. An observer can become what is observed. And *vice versa*. In this sense, all physical systems are on the par, and there is no privileged observer. However, at the same time, – and this is what Rovelli does not take into account – once in a context an observer and an observed system are determined, there always is a categorical difference between them.

We agree with Rovelli that there is no need to introduce "subjective states of consciousness", since any physical system can play the role of the observer. He writes: "As soon as we relinquish this exception, and realise that any physical system can play the role of a Copenhagen's "observer", we fall into relational quantum mechanics. Relational quantum mechanics is Copenhagen quantum mechanics made democratic by bringing all systems onto the same footing." [2]

Like Rovelli, we go back to the Copenhagen interpretation and claim that it must be correctly understood, corrected. (Later metaphysical interpretations turned out to be worse than the original interpretation of the founding fathers of quantum physics (although, notice, they had different views, and the term "Copenhagen interpretation" characterises the state of affairs only approximatively).) In doing so and unlike Rovelli, we accept the categorical distinction between the ideal observer (theory, norm, concepts) and the real (observed) system (application of a theory, real objects). This allows one to avoid the metaphysical problems encountered by various interpretations of quantum mechanics, including the relational one and the so-called QBism.[4]

## 5. Quantum Phenomenon and Quantum Observer

Our return to the Copenhagen interpretation, corrected in the framework of contextual realism, can be compared with a return to the logic of a phenomenon.

The notion of a phenomenon was introduced by Plato. And he discovered that a phenomenon has a normative structure: it presupposes the distinction between appearance and reality. An appearance can correspond or not to the reality of things. In other words, a phenomenon presupposes a judgment, made by a "classical" subject according to a norm. An appearance corresponds to reality if and only if the corresponding judgment is correct. Aristotle developed the notion of a phenomenon. For him, the phenomenon has the following logic: "(…) the appearance is true; not in itself, but for him to whom it appears, and at, the time when it appears, and in the way and manner in which it appears" [15].

As Jocelyn Benoist has shown, in the history of philosophy, the notion of a phenomenon was deformed and truncated, and (in the phenomenology of the 20[th] century) absolutized and naturalized. For example, for Kant, as well as for Plato, a phenomenon is something that is not in the object itself, but appears in relation to the subject. The Kantian phenomenon has a cause. However, this cause is the unknowable thing-in-itself. Benoist characterizes the Kantian phenomenon as a "handless phenomenon". August Comte's phenomenon does not have any cause at all. In Benoist's words, this is a "phenomenon-orphan". And so on. [16]

A correct understanding of the notion of a "quantum phenomenon" requires a return to the correct Platonic-Aristotelian notion of a "phenomenon". In our Wittgensteinian terminology, a phenomenon is a "language game", governed by a norm/rule, within which a real object is identified [4]. A quantum phenomenon is the identification of a quantum "object" in a broad sense (for example, a quantum correlation)[5] with the help of a quantum theory used by a classical subject in the classical space-time. The preparation of an experimental situation is part of the application of a theory.

---

4 According to QBism – a variety of quantum Bayesianism – Born's rule and quantum theory have the normative, not the descriptive, character; they do not describe the objective reality, existing independently from the subject and the language use – the ordinary as well as the theoretical [13; 14; 21]. In this lies some similarity with our approach, according to which the quantum theory is a Wittgensteinian norm (rule) [6]. However, QBism considers the quantum measurement as an interaction of the subject with a quantum system, allowing to the former to affect the latter. So, for QBism, only the result of an interaction is measured and not the real things as they actually exist independently from the subject. This is rather an idealist, not a realist approach.

5 Quantum objectivity does not need the classical objects.

According to Bohr's neo-Kantian view, the use of classical terminology in the description of quantum experiments is unavoidable because such a description supposes a description of the measurement instrument, including its position in space and functioning in time [17, 18]. Our non-metaphysical realist's view does not contradict Bohr's view and surpasses it. We treat Bohr's notion of a quantum experience ("experiment"), in which the observer participates, as a quantum phenomenon – an application of a theory to reality, a language game. The quantum "observer" is not a subjective consciousness; it belongs to the logic of a phenomenon [19].

## 6. Conclusions

Quantum mechanics and the law of quantum probability (the Born rule) do not describe an autonomous determinate reality, which would be independent from them in an absolute sense. The correspondence between them can be compared with the correspondence between cooking recipes and food prepared according to these recipes. (Wittgenstein uses the metaphor of cooking recipes to indicate the grammatical difference between the purely instrumental rules, like chess rules, and, let us say, the real rules, that is, those anchored in reality [20].)

So, a wave function neither literally reflects an "exterior reality", nor is it merely a predictive mathematical tool. In particular, an entangled wave function can be viewed as the formal cause of quantum correlations whose reality is contextual. The reduction of a wave function is not a real physical process. And there are no autonomous quantum events which would be different from the facts of their representation. Ontology is contextual [22]. It is wrong to think that while measuring a physical quantity something happens independently of the formalism of quantum mechanics and only later it is expressed in the formalism. Independently of the means of identification, nothing happens, appears. Thus, the classical dualism of the event and the fact (describing the event) is rejected.

To solve (or dissolve) the metaphysical and conceptual problems of quantum mechanics, it is necessary to take into consideration the logical difference – the logical "gap" – between the rule (theory, evolution of a wave function) and its real use (application of a theory). The quantum measurement problem (as well as other metaphysical problems of quantum mechanics) results from a confusion between the categories of the ideal and the real.

frigpr@gmail.com